\documentclass[12pt]{article}

\usepackage{epsfig}
                        
\let\section=\subsection     \let\subsection=\subsubsection                

\newcommand{\gapprox}{{\raisebox{-0.5ex}{${\scriptstyle>}$} \atop \raisebox{0.5ex}{${\scriptstyle\sim}$}}}
\newcommand{\lapprox}{{\raisebox{-0.5ex}{${\scriptstyle<}$} \atop \raisebox{0.5ex}{${\scriptstyle\sim}$}}}

\begin{document}

\begin{center}
{\large \bf Medium Effects and the Structure of}\\[2mm]
{\large \bf Neutron Stars in the Effective Mass Bag Model
\footnote{Supported by BMBF, GSI Darmstadt, and DFG}}\\[5mm]
K.~SCHERTLER\footnote{E-mail: klaus.schertler@theo.physik.uni-giessen.de}, C.~GREINER, and M.~H.~THOMA{
\footnote{Heisenberg Fellow}} \\[5mm]
{\small \it Institut f\"ur Theoretische Physik, Universit\"at Giessen \\
35392 Giessen, Germany \\[8mm] }
\end{center}

\begin{abstract}
\noindent
One of the most intriguing consequence of the extreme conditions inside 
neutron stars is the possibility of the natural existence of a deconfined 
strange quark matter phase in the high density interior of the star.
The equation of state (EOS) of strange quark matter (SQM) was recently 
improved in the framework of the MIT bag model by including medium effects. 
It was found that medium effects increase the energy per baryon
of SQM and therefore lower the stability of this phase. 
In this work we investigate the influence of medium effects on the structure 
of hybrid stars within this model.  
We found that the medium effects reduce the extent of a pure SQM phase 
in the interior of an hybrid star significantly in favor of a mixed phase
of quark and hadronic matter.
\end{abstract}

\section{Introduction}
The possibility of a deconfined phase of strange quark matter (SQM) in the 
interior of neutron stars is still stimulating the work of many authors 
\cite{Glen92,Hybrid}. 
The gross structure of a neutron star like its mass and radius (MR) is 
influenced by the composition of its stellar material. 
This holds especially in the case of the existence of strangeness bearing 
``exotic'' components like hyperons, kaons or SQM which may significant change the 
characteristic MR relation of the star.  

The scope of this work is to study the influence of medium effects in the SQM
phase on the gross structure of hybrid stars, i.e. neutron stars which are made 
of hadronic matter in the outer region, but with a SQM core in the interior. 
The deconfinement phase transition from hadronic matter to the SQM phase 
is constructed according to Glendenning \cite{Glen92}. 
We only require the weaker condition of global charge neutrality instead of 
assuming charge neutrality in either phase. The latter assumption would have the 
drastic consequence of strictly excluding a possible mixed phase of quark matter 
and hadronic matter. 
Such mixed phase is supposed to form a crystalline lattice of various geometries 
of the rarer phase immersed in the dominant one and probably exists over a wide
range of densities inside the star \cite{Glen92, Hybrid}.

\section{SQM and the effective mass bag model}

SQM has been suggested as a possible stable or metastable 
phase of nuclear matter \cite{Bodm71Witte84}.
The equation of state of this system is commonly described as a non-interacting 
Fermi gas of quarks at zero temperature, taking into account the bag constant 
\cite{Bodm71Witte84,FahrJaff84FreeMcLe78}. Also quark interactions within lowest order 
perturbative QCD have been considered.

In condensed matter as well as in nuclear physics medium effects play an 
important role. One of the most important medium effects are effective
masses generated by the interaction of the particles with the system.
The EOS of SQM was recently improved in the framework of the MIT bag model 
by including these medium effects \cite{Sche97}. 
The quarks are considered there as quasi-particles which acquire an effective 
mass by the interaction with the other quarks of the dense system. 
The effective masses following from the hard dense loop quark self energy 
are given in \cite{Sche97, EffeMass}.
They are used in the ideal Fermi gas EOS at temperature $T=0$ with respect to thermodynamic
self-consistency in the sense of \cite{GoreYang95}. 
It was found \cite{Sche97}, that the energy per
baryon of SQM increases with increasing coupling constant $g$ which enters into the effective masses. 
This makes the SQM phase energetically less favorable.
We will refer to this model \cite{Sche97} as the ``effective mass bag model''.

\section{The gross structure of neutron stars}

Now we want to use the EOS in the effective mass bag model to calculate the MR relation of pure SQM stars 
and hybrid stars by solving the Tolman-Oppenheimer-Volkoff equations \cite{OppeVolk39}. 
For the SQM EOS we assume a bag constant of $B^{1/4}=165$ MeV ($B\approx 96$ MeV/fm$^3$) and current quark 
masses of $m_u=m_d=0$, $m_s=150$ MeV \cite{ScheHybstar}. For the hadronic phase
of the hybrid star we use an EOS calculated in the framework of the nonlinear Walecka model including nucleons, 
hyperons ($\Lambda$ and $\Sigma^-$), electrons, muons and $\sigma$, $\omega$ and $\rho$ mesons \cite{ScheHybstar, RMF}. 
We choose a compression modulus of $K=300$ MeV. For subnuclear densities we use the Baym-Pethick-Sutherland 
EOS \cite{BPS}.
%
\begin{figure}[ht]
\centerline{\epsfig{file=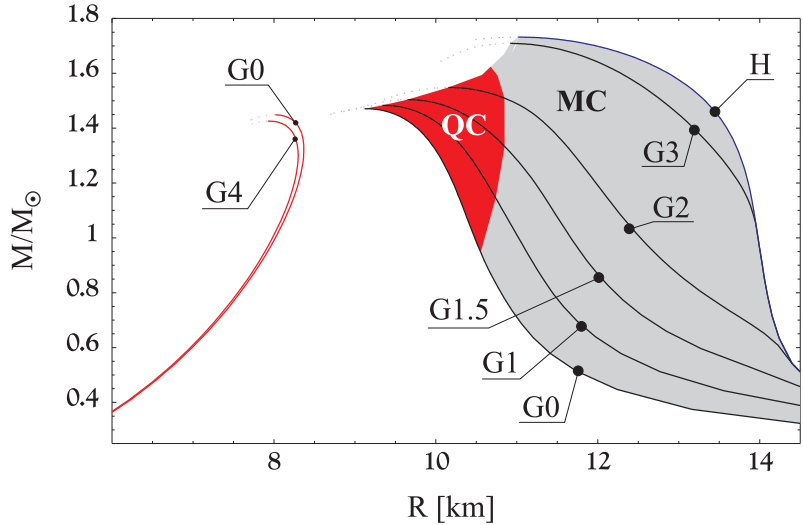,height=7cm}}
\caption{Mass radius relation for pure SQM stars ($R<9$ km) and hybrid stars ($R>9$ km), 
G0 = (g=0), \dots, H = pure hadron
, QC = star has a quark core, MC = star has a mixed core, $B^{1/4}=165$ MeV,
$K=300$ MeV.}
\label{RM1}
\end{figure}
%
Fig.\,\ref{RM1} shows the resulting MR relations for various values of the coupling constant $g$ \cite{ScheHybstar}. To investigate
the influence of medium effects on the MR relation, we consider $g$ as an parameter ranging from $g=0$ 
(no medium effects) to $g=4$ \cite{ScheHybstar, PeshScheThom97}.
The left hand side ($R<9$ km) shows the pure SQM star results. As already found in \cite{Sche97},
we see that medium effects have only an slight influence on the MR relation of a pure SQM star. This situation 
changes if we look at the hybrid star results on the right hand side of fig.\,\ref{RM1} ($R>9$ km). 
With increasing $g$, the MR relation approaches the curve of the pure hadron star (denoted by H). There are 
two different shaded regions denoted by QC (quark core) and MC (mixed core). Every star located inside the 
QC region possess a pure SQM core while the MC region denotes stars with a mixed phase core.

To discuss the radii of the quark and mixed cores, we assume an canonical mass of \mbox{$M=1.4 M_\odot$}.
Fig.\,\ref{R} shows the schematic view of the canonical star for different increasing $g$ \cite{ScheHybstar}. We find 
that a small coupling constant of $g=1.5$ ($\alpha_s\approx0.18$) is able to shrink the radius of the pure quark phase (QP) from 
$R \approx 6$ km (with neglected medium effects, fig.\,\ref{R}a) to $R \approx 3$ km (fig.\,\ref{R}c). 
Already at $g=2$ ($\alpha_s\approx0.32$) the pure SQM core is vanished completely (fig.\,\ref{R}d).
Note that in spite of a completely vanishing quark core, the pure hadron phase (HP) 
has grown only moderately. One could say that in a wide range of $g$ ($g\lapprox3$) medium effects are not able to
displace the quark phase in favor of a pure hadronic phase. The essential effect is the transformation of the
pure SQM phase into a SQM phase immersed into the mixed phase (MP) which therefore dominates the star.
Only for $g\gapprox3.5$ a phase transition to SQM is completely suppressed (fig.~\ref{R}f).
%
\begin{figure}[ht]
\vspace{3cm}
\[\mbox{\tt figure2.gif}\]
\vspace{3cm}
\caption{Schematic gross structure of an $M=1.4 M_\odot$ star.}
\label{R}
\end{figure}
%

\section{Conclusion}
We have investigated the gross structure of non-rotating pure strange quark matter and hybrid stars using the effective 
mass bag model \cite{Sche97, ScheHybstar} for the description of the quark matter EOS. 
This model is based on the quasi-particle picture where
the quarks of a Fermi-gas acquire medium-dependent effective quark masses generated by the interaction of the quarks with the other
quarks of the system.
We found that the basic influence of medium effects described by this model and parameterized by the
strong coupling constant $g$ is to reduce the extent of a pure quark matter phase in the interior of an hybrid star 
significantly in favor of a mixed phase. 
For a wide range of the coupling constant ($g\lapprox3$, $\alpha_s\lapprox0.72$) 
SQM is therefore present in the dense interior of the star at least as a mixed phase of quark and hadronic matter.

\bigskip
{\bf Acknowledgements}

We like to thank P.K. Sahu for providing us with the hadronic EOS.

\end{document}